\documentstyle[epsfig]{aipproc}

\begin{document}
\title{Properties of Gamma-Ray Burst Classes}

\author{Jon Hakkila$^*$, David J. Haglin$^*$, Richard J. Roiger$^*$,\\
Robert S. Mallozzi$^{\dagger}$, Geoffrey N. Pendleton$^{\dagger}$, \&
Charles A. Meegan$^{\ddagger}$}
\address{$^*$Minnesota State University, Mankato, Minnesota 56001\\
$^{\dagger}$University of Alabama, Huntsville, Alabama 35812\\
$^{\ddagger}$NASA/MSFC, Huntsville, Alabama 35812}

\maketitle

\begin{abstract}
The three gamma-ray burst (GRB) classes identified by statistical clustering 
analysis \cite{mukherjee98} are examined using the pattern recognition 
algorithm C4.5 \cite{quinlan86}. Although the statistical existence of 
Class 3 (intermediate duration, intermediate fluence, soft) is supported, 
the properties of this class do not need to arise from a distinct source 
population. Class 3 properties can easily be produced from Class 1 (long, 
high fluence, intermediate hardness) by a combination of measurement error, hardness/intensity correlation, and a newly-identified BATSE bias 
(the fluence duration bias). Class 2 (short, low fluence, hard) 
does not appear to be related to Class 1.
\end{abstract}

\section*{Introduction}
GRB spectral and temporal properties overlap, providing a continuum of 
burst characteristics. Some of this overlap is intrinsic in nature, while 
much is due to instrumental and observational biases. In addition to this overlap, there is clustering indicative of classes within the parameter 
space defined by GRB attributes. In particular, there are two 
long-recognized GRB classes \cite{cline74,kouveliotou93} based on 
duration (divided at roughly 2 seconds) and spectral hardness. A 
statistically significant third class has been identified using 
statistical clustering analysis \cite{mukherjee98}. 

Can effects attributable to a source population 
be separated from instrumental effects? To answer this, we have applied 
computer science pattern recognition algorithms to learn why bursts 
cluster in some parameter spaces. For this analysis, we have used the 
supervised decision tree classifier C4.5 \cite{quinlan86}. Supervised 
classifiers establish rules for previously identified patterns, and 
must be trained by representative class members. 
 
\section*{Analysis}

The three GRB classes identified by statistical clustering techniques 
\cite{mukherjee98} can be found from three significant classification attributes; 50 to 300 keV fluence, T90 duration, and HR321 hardness 
ratio (the fluence in the 100 to 300 keV band divided by the fluence 
in the 25 to 100 keV band). The properties of the three classes in 
terms of these attributes are demonstrated in Table \ref{table1}.

\begin{table}[ht!]
\caption{Statistical clustering classes, from 3B GRBs.}

\label{table1}
\begin{tabular}{lccc}
 Attributes& Class 1 (Long)& Class 2 (Short)& Class 3 (Intermediate) \\
\tableline
T90: & long & short & intermediate \\
Fluence: & large & small & intermediate \\
Hardness: & intermediate & hard & soft \\
\end{tabular}
\end{table}

C4.5 was trained on the three GRB classes using five fluences, two 
durations, three peak fluxes, and three hardness ratios. 
C4.5 produced a decision tree containing IF THEN ELSE branches for placing 
each GRB in the appropriate class; 
these branches were {\it pruned} to remove branches containing less than 
four GRBs. Rules were then generated for each class based on the pruned branches. C4.5 identifies outliers with poorly defined rules that often 
contain few GRBs. Statistical methods find that outliers are not closely 
bound to the class (cluster) centers.
C4.5 rules identified a number of GRBs as having peculiar hardness 
ratios; these resulted from large individual channel fluence errors. 
The GRBs with the largest 10\% relative errors (error divided by 
measurement) were subsequently removed from the database. The remaining 3B 
GRBs were reclassified using C4.5; the resulting rules were used to 
classify 4B Catalog GRBs and thus increase the database size. 

\subsection*{Class 3 Spectral Hardnesses}

C4.5 verified that the three GRB classes resulted primarily from the attributes of spectral hardness, duration, and fluence. With the larger classification database, the dependence on spectral hardness could be examined in terms of the spectral fitting parameters $\alpha$, $\beta$, and E$_{\rm peak}$ \cite{band93}.
Using only these three attributes, C4.5 was able to accurately classify most of the 4B GRBs. The rules generated by C4.5 were able to cleanly separate Class 2 from Class 1, but could not delineate Class 3 from Class 1 (85\% of Class 3 bursts were assigned to Class 1).

\begin{figure}[ht!] 
\centerline{\epsfig{file=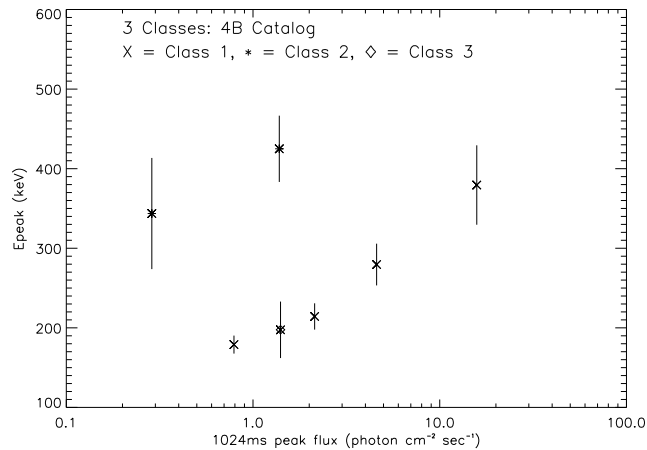,width=3.5in}}
\vspace{10pt}
\caption{E$_{\rm peak}$ vs. p1024 for the Three GRB Classes.}
\label{fig1}
\end{figure}

Upon further examination, Class 3 GRBs were found to have E$_{\rm peak}$ 
values similar to Class 1 bursts of the same 1024 ms peak flux 
(Figure \ref{fig1}). The correlation between E$_{\rm peak}$ and 
peak flux has been interpreted as cosmological redshift \cite{mallozzi95}.

\subsection*{Class 3 Fluences and Durations}

Since at least one of the three defining characteristics of Class 3 
actually represents a data correlation, we hypothesized that Class 3 
GRBs actually belong to Class 1. We decided to see if Class 3 fluences 
and durations could be explained in terms of Class 1 attributes. This 
could be the case if some instrumental or sampling bias made Class 1 
GRBs appear to be shorter and fainter than they should be.

\begin{figure}[ht!] 
\centerline{\epsfig{file=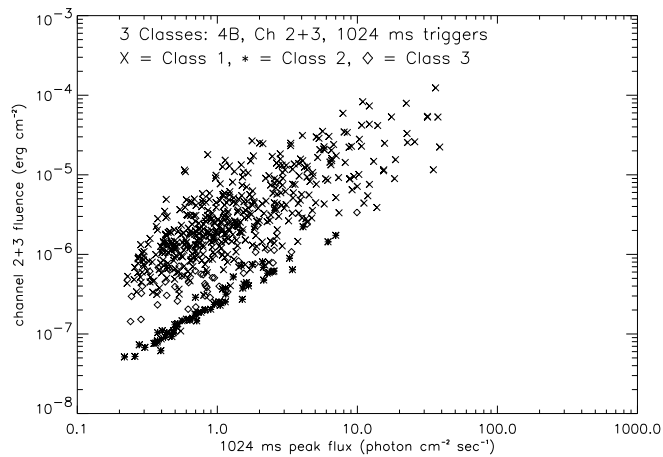,width=3.5in}}
\vspace{10pt}
\caption{Fluence vs. p1024 for the Three GRB Classes.}
\label{fig2}
\end{figure}

Figure \ref{fig2} is a plot of fluence vs. 1024 ms peak flux for each 
of the three GRB classes, and is limited to GRBs detected when BATSE 
had one homogeneous set of trigger criteria. 
There are distinct regions outside of which no GRBs are 
found. GRBs with 1024 ms peak fluxes less than 0.2 photons cm$^{-2}$ 
second$^{-1}$ are not detected, since this is below BATSE's minimum 
detection threshold. GRBs do not have fluences less than what would be 
found in their time-integrated 1024 ms peak fluxes, since this is the 
shortest timescale on which this peak flux can be measured. 

\begin{figure}[ht!] 
\centerline{\epsfig{file=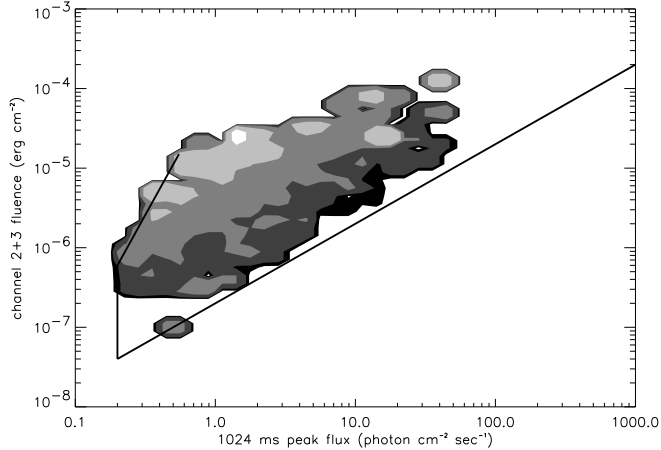,width=3.5in}}
\vspace{10pt}
\caption{Fluence vs. p1024 for Class 1 GRBs; contours indicate regions of constant log(T90).}
\label{fig3}
\end{figure}

Figure \ref{fig3} overlays $\log$(T90) contours for Class 1 GRBs on the 
fluence vs. 1024 ms peak flux space. The contours demonstrate that 
GRBs can be modeled as a series of pulses, with pulses containing most of 
the fluence and interpulse separations primarily defining the duration. 
Most Class 2 bursts are single-pulsed events as measured on the 1024 
ms timescale.  This helps define the characteristics of the third 
distinct region outside of which no GRBs are found: high fluence, 
faint Class 1 GRBs are missing, whereas low fluence faint, Class 1 
GRBs are present. Since a bias favoring detection of GRBs with few 
photons over those with many photons seems unlikely, we suspect a bias 
capable of underestimating fluence relative to peak flux.

We have dimmed a number of bright GRBs to where they just trigger 
in order to study their measured properties as they fade into 
background. Each burst's peak flux is dimmed, and the time history 
is ``noisified'' with a Poisson background. The peak flux and fluence 
are then re-measured. These actions have been performed ten times on five 
bright bursts with a range of temporal structures.

\begin{figure}[ht!] 
\centerline{\epsfig{file=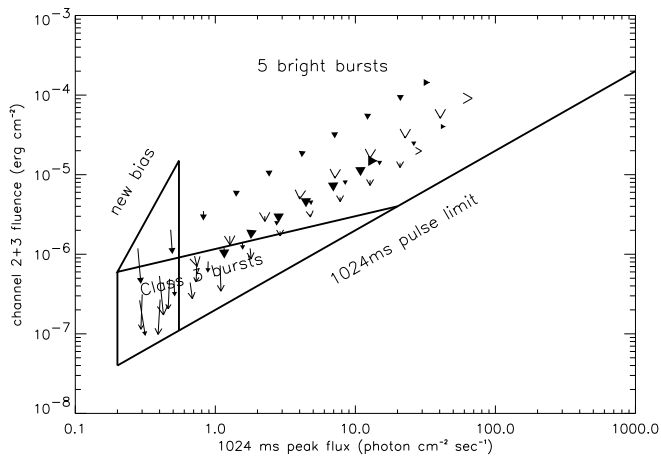,width=3.5in}}
\vspace{10pt}
\caption{Five bright Class 1 GRBs, decremented in peak flux, noisified, with remeasured fluences and peak fluxes. It has been assumed that the GRB duration is measured from identifiable pulses, which become harder to recognize as the peak flux becomes fainter.}
\label{fig4}
\end{figure}

One problem quickly became apparent during the analysis: the time 
interval bounding the fluence measurement (the {\it fluence duration} 
\cite{hakkila99}) strongly influenced the amount of fluence measured. 
If the same fluence duration interval was used for undimmed and dimmed 
measurements, then the fluence-to-peak flux ratio did not change as a GRB 
was dimmed. If, however, the fluence duration interval shortened to 
account for faint pulses disappearing into the background 
and becoming unrecognizable, then the fluence-to-peak flux ratio 
decreased as the burst dimmed (see Figure \ref{fig4}). This bias becomes
stronger near the trigger threshold.

Fluence durations taken from BATSE Catalogs provide supportive evidence 
for this mechanism. The durations used to calculate fluence of faint 
Class 1 GRBs are shorter than those of bright Class 1 GRBs \cite{hakkila99}. 

\section*{Conclusions}

A mechanism exists whereby some Class 1 (Long) GRBs can develop 
Class 3 (Intermediate) characteristics via a combination of the 
hardness intensity relation and the fluence duration bias. Faint Class 
1 GRBs are most likely to develop Class 3 characteristics, but it is 
possible for even bright GRBs with appropriate time histories and 
spectral features to develop these characteristics. Class 3 
(Intermediate) GRBs do not therefore appear to represent a separate 
source population, although they cluster in the duration, fluence, 
hardness, attribute space. Class 2 (Short) GRBs do appear to represent 
a separate source population. 
We were unable to find a mechanism by which faint Class 1 GRBs could 
develop Class 2 characteristics.

GRB population studies can benefit from use of AI classifiers. There are 
many other attributes developed by the community that could be included 
for future study. To this end, we are designing a web-based AI tool for 
GRB classification \cite{haglin99} that includes supervised and 
unsupervised AI classifiers \cite{roiger99}.

\end{document}